\documentclass[12pt]{iopart}

\usepackage{graphicx}

\begin{document}

\title
[Time-resolved studies on the collapse of Mg atom foam in He droplets] 
{Time-resolved studies on the collapse of magnesium atom foam in helium nanodroplets}

\author{S. G\"ode, R. Irsig, J. Tiggesb\"aumker, K.-H. Meiwes-Broer}

\address{Institut f\"ur Physik, Universit\"at Rostock,
  Universit\"atsplatz 3, 18051 Rostock, Germany}
\ead{josef.tiggesbaeumker@uni-rostock.de}

\begin{abstract}
  Magnesium atoms embedded in superfluid helium nanodroplets have been
  identified to arrange themselves in a metastable network, refered to as foam. In order to investigate the ionization dynamics of this unique
  structure with respect to a possible light-induced collapse the femtosecond dual-pulse spectroscopy technique is applied. Around
  zero optical delay a strong feature is obtained which represents a
  direct probe of the foam response. We found that upon collapse,
  ionization is reduced. A particlar intensity ratio of the pulses
  allows to address either direct ionization or photoactivation of the neutral complexes, thus affecting reaction pathways. A simplified excitation scheme visualizes possible scenarios in accordance with the experimental observations.
  
\end{abstract}

\pacs{
  36.40.-c, %(Atomic and molecular clusters (see also 61.46.-w
                %Nanoscale materials in condensed matter)),
  31.70.Hq, %(Time-dependent phenomena: excitation and relaxation
            %processes, and reaction rates (for chemical kinetics
            %aspects, see 82.20.Rp)),
  71.30.+h, %(Metal-insulator transitions and other electronic
            %transitions),
%  33.20.Xx, %(Spectra induced by strong-field or attosecond laser
            %irradiation ),
  33.80.Rv %(Multiphoton ionization and excitation to highly excited
           %states (e.g., Rydberg states) }
}

%\maketitle
\section{Introduction}

In cluster physics of special interest is, the size-dependent
evolution of fundamental properties~\cite{JorZPD92}, e.g.\ changes in bond character, in particular the non-metal to metal transition phenomenon~\cite{Red10}. Much effort has been devoted to identify the critical size N$_{\rm c}$ characterizing the transitional regime, see
e.g.\ Ref.~\cite{IssARPC05} for a summary. For magnesium, which is the
subject of this study, it was found that the transition takes place
at around to N$_{\rm c}$=20 atoms~\cite{DiePRL01,ThoPRL02}. Meanwhile it became
possible to determine N$_{\rm c}$ also in elements with s$^2$p$^2$
orbital configurations like tin~\cite{CuiJCP07} or lead~\cite{SenPRL09}. Up until now, studies on clusters have concentrated on various aspects that depend on the number of their constituents. Hence, refering to the original suggestion by Mott~\cite{mott} to induce a phase transition via the change of the atomic density has not been tackled with clusters up to now. In general, when small clusters in the gas phase aggregate, they do so near their electronic and structural ground state with excitations on the order of k$_{\rm B}$T. An interesting issue would be to study finite systems in which the interatomic distances can to some extend be controlled. Examples are atoms in optical lattices~\cite{BloRMP08} and cold Rydberg gases where the van der Waals interaction leads to Coulomb blockade of optical transitions~\cite{LukPRL01}. However, these systems are designed to probe the response at $\mu$m interatomic distances. On the nm-scale, aggregation in superfluid helium droplets may lead to structural~\cite{NauS99} and electronically metastable~\cite{HigS96,SchPRL04,PrzPRB06} %,NagPRL08}
configurations, offering a unique possibility to study matter at conditions far from the ground state. 

Recently, we found evidence that Mg atoms might arrange in a unique
foam-like assembly~\cite{PrzPRA08}. Those experiments have been conducted by means of two-photon ionization mass spectrometry with nanosecond laser pulses, taking the atomic \mbox{$3~^1P_1^0 \leftarrow 3~^1S_0$} transition of single embedded Mg atoms as a spectral reference~\cite{RehJCP00a,MorJMS06}. For droplets containing more than one atom on average, a nearby narrow peak
shows up that is red-shifted by a few nm, i.e.\ from $\lambda_{\rm
  atom}$=279\,nm to $\lambda_{\rm foam}$=282\,nm. The optical spectra of \emph{all} clusters (up to Mg$_{14}^+$), which appear as result of the excitation in the mass spectra, peak at the same $\lambda_{\rm foam}$. This spectroscopic finding suggests that before excitation single Mg atoms are dissolved within the droplet. From the peak shift with respect to the monomer signal, an interatomic Mg-Mg distance of 10\,\AA\, has been deduced. It is therefore likely that the atoms arrange in a regular foam-like network. We can thus assume, that the increase in helium density around Mg impurities as a result of van der Waals forces and modified by the superfluid properties of the nanodroplet~\cite{DalZPD94} leads to the formation of a shallow potential barrier on the order of a few K. This effect prevents Mg atoms from collapsing to the cluster ground state. Our results are supported by chemi-luminescence studies on Mg atoms embedded in He droplets conducted in the Huisken group~\cite{KraJPCA10}. From the theoretical side density functional calculations by the group of Barranco confirm that a helium-induced potential barrier prevents the formation of ground state Mg dimers~\cite{HerPRB08,HerJPCS09}. Such a foam-like structure has also been predicted for embedded neon atoms~\cite{EloPRB08}. In those simulations, a network with an average Ne-Ne spacing of about 6\,\AA\ was found, which the author called \emph{quantum gel}. We would like to point out that the foam state is different from impurity helium solids, where compact clusters or nanoparticles are encapsulated by solidified helium~\cite{GorJLTP04}.

Because the barrier is expected to be low, photoabsorption will be sufficient to destroy the metastable foam state and initiate a rapid collapse on a ps timescale. In the present contribution we follow this idea using femtosecond pump-probe spectroscopy to trigger the transition and study the temporal development of the involved dynamics. Dual fs pulses with a given intensity ratio have been applied to explore the relaxation pathways. Beside the transformation into compact clusters, exciplexes and ion-snowballs can be formed, whereby the yields crucially depend on the pulse order. The analysis provides further information on the complex time-evolution of the activated foam.

\section{Experimental Setup}

\begin{figure}[t]
\centering
\vspace{0.5cm}
\includegraphics[width=14cm]{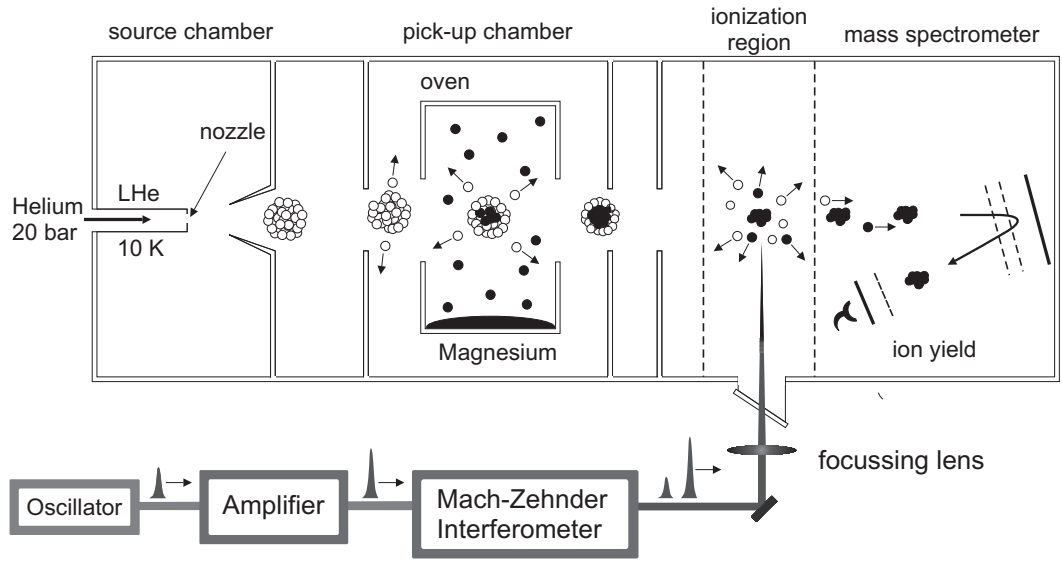}
\caption{Schematic view of the experimental setup to study Mg
  complexes embedded in helium nanodroplets by femtosecond dual-pulse
  spectroscopy. An attenuator in one of the interferometer arms allows
  to select the pulse intensity ratio. The products of the interactions are      analyzed by time-of-flight mass spectrometry.}
\label{fig:ExpSetup}
\end{figure}

A schematic view of the experimental setup is shown in
\mbox{Fig.~\ref{fig:ExpSetup}}. The femtosecond laser system provides
pulses with a variable width (30\,fs -- 1\,ps) at a repetition rate of
1\,kHz with a maximum energy of 2.5\,mJ at $\lambda_0$= 810\,nm.  The
pulses are fully analyzed using a home-built SHG-FROG
device~\cite{TruMST10}. In the experiment we choose a pulse width of
about 200\,fs (FWHM) to avoid strong nonlinearities. For the dual-pulse
measurements, the optical delay is generated by a Mach-Zehnder
setup. An additional attenuator in one of the interferometer arms
allows to select the pulse energy ratio.

The helium droplet pick-up technique~\cite{SchPRL90,GoyPRL92} is
applied to produce complexes of embedded Mg atoms, details of which are
provided elsewhere~\cite{ToeAngChemIntEd04,StiJPB06,TigPCCP07}. In short, cold helium gas (20\,bar) is expanded through a 5\,$\mu$m orifice. By tuning
the nozzle temperature between 9--14\,K, the droplet sizes vary
from 70 to 7\,nm~\cite{BucJCP90}. The molecular beam passes through a 4\,cm
long heated cell containing magnesium pellets. By tuning the oven
temperature, the number of atoms being picked up by the droplets can
be adjusted. After differential pumping the doped droplets enter the
interaction region where the laser beam perpendicularly
intersects the molecular beam. The intensity conditions have carefully
been checked and calibrated to the ion appearance intensity of xenon
gas~\cite{AugPRL89} with an uncertainty factor of about two. The
resulting ionic products of the interactions are analyzed by
reflectron time-of-flight mass spectrometry (TOF).

In the present study the source is tuned to conditions similar
to those in Ref.~\cite{PrzPRA08}: helium droplets with a mean size of about 40,000 atoms are doped with $\rm N_{\rm avg}$=7 Mg atoms on average. We
cross-check the actual target composition by simulating the pick-up
process taking the helium droplet distribution for the experimental
nozzle temperature and Mg vapor pressure in the oven cell into account
(T$_{\rm source}$=10K, p$_{\rm Mg}$=$5\times10^{-5}$mbar). A
slightly off-focus configuration of the laser beam with respect to the
molecular beam axis leads to a power density of \mbox{$\rm I=6\times10^{11}\,W\,cm^{-2}$}. At this reduced laser intensity Mg atoms undergo multiphoton ionization (MPI) whereas charging of helium is negligible due to its higher ionization energy. Non-resonant laser conditions are choosen, i.e., neither $\lambda_0$ nor its harmonics are close to the foam resonance at $\lambda_{\rm foam}$=282\,nm~\cite{PrzPRA08}. However, resonant enhanced photoionization via highly excited electronic states might be possible.

\section{Results}

\begin{figure}[t]
  \centering
  \vspace{0.5cm}
  \includegraphics[width=11cm]{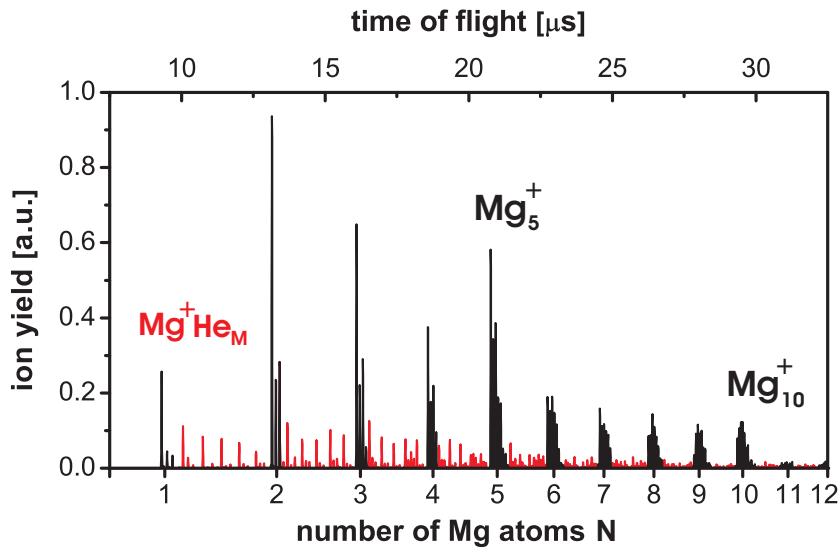}
  \caption{Typical time-of-flight spectrum of magnesium complexes
    (N$_{\rm{avg}}$=7) formed in 10\,nm-sized helium droplets (40.000
    atoms) recorded after femtosecond multiphoton ionization. As a
    result of the interaction, Mg$^+_{\rm N}$ clusters and
    Mg$^+$He$_{\rm M}$ snowballs (red) are formed. The single pulse
    laser intensity is $6\times10^{11}$\,W\,cm$^{-2}$. Note, that the              multiple-peak structure of Mg$^+_{\rm N}$ arises from the atomic isotope distribution.}
  \label{fig:massSpectra}
\end{figure}

\mbox{Fig.~\ref{fig:massSpectra}} depicts a section of a TOF spectrum resulting from ionization with fs laser pulses. Cluster ions with N$\le$20 atoms can be detected. Concentrating on the low mass range, Mg$^{+}_{\rm 5}$ and Mg$^{+}_{\rm 10}$ show up with increased yields, indicating that these are more stable compared to adjacent ones. Similar features have also been observed utilizing ns-excitation as well as electron impact~\cite{DiePRL01}. The spectrum shows strong contributions from ion-snowballs (Mg$^+$He$_{\rm M}$, marked red in the figure), a well-known product observed in strong-field ionization, such as in Coloumb explosion of doped droplets~\cite{DoePCCP07}. From a single mass spectrum alone the dynamics of the foam cannot be deduced. In order to identify the foam response one has to control the initial excitation, followed by an interrogation step. We manage this by fs-dual-pulse measurements choosing a pulse energy ratio of \mbox{E$_{\rm s}$/E$_{\rm w}$=3}, with peak intensities in the interaction volume of \mbox{I$_{\rm s}$=$6\times10^{11}$\,W\,cm$^{-2}$} for the strong and \mbox{I$_{\rm w}$=$2\times10^{11}$\,W\,cm$^{-2}$} for the weak pulses, respectively. At these conditions a ten times higher ionization rate is observed for I$_{\rm s}$ (total rate at 1\,kHz repetition rate \mbox{Y$^{\rm tot}_{\rm s}$=210\,s$^{-1}$}) compared to I$_{\rm w}$ (\mbox{Y$^{\rm tot}_{\rm w}$=18 s$^{-1}$}). Hence, the strong initial pulses create charged foam complexes while the weak ones predominantly yield neutral excited (\emph{photoactivated}) systems. In the experimental runs the time delays of the dual-pulses as well as their order are varied.

\begin{figure}[t]
  \centering
	\vspace{0.5cm}
  \includegraphics[width=9cm]{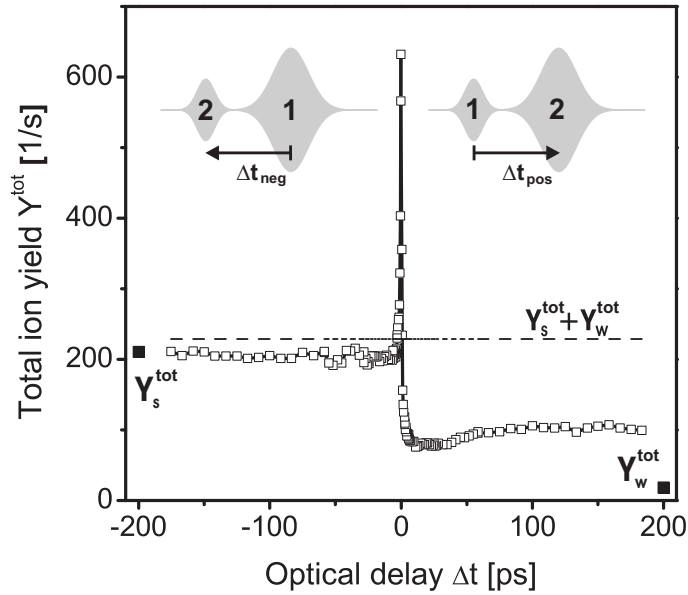}
  \caption{Total ion yields after dual-pulse excitation of the embedded Mg foam. Evidently, the signal sensitively depends on whether the strong pulse is applied first or second. If the ionizing pulse arrives after the weak initial pulse ($\Delta$t$_{\rm pos}$) the yield is reduced by a factor of two compared to the reversed order ($\Delta$t$_{\rm neg}$). The strong peak around zero delay hints at an ultrafast response of the excited aggregates and contains a small autocorrelation contribution. Yields obtained for \emph{single} strong (Y$^{\rm tot}_{\rm s}$) and \emph{single} weak (Y$^{\rm tot}_{\rm w}$) excitations are indicated by bold symbols. The dashed line represents the sum of these signals \mbox{(Y$^{\rm tot}_{\rm s}$+Y$^{\rm tot}_{\rm w}$)}. Laser intensities: \mbox{I$_{\rm s}$=$6\times10^{11}$\,W\,cm$^{-2}$} and \mbox{I$_{\rm w}$=$2\times10^{11}$\,W\,cm$^{-2}$}.}
  \label{fig:pptotalYield}
\end{figure}

\mbox{Fig.~\ref{fig:pptotalYield}} shows the dual-pulse total ion yield \mbox{Y$^{\rm tot}$} as function of pulse separation $\Delta$t. A distinct dependence on the order of the pulses is obtained. Moreover, exept for the range around zero delay the dual-pulse signal stays well below that of the sum of individually applied pulses (see the dashed line in \mbox{Fig.~\ref{fig:pptotalYield}}). This observation is less pronounced with strong leading pulses, i.e. negative delays $\Delta$t$_{\rm neg}$, but becomes distinct when ionization with the strong pulse is delayed, i.e. positive delays $\Delta$t$_{\rm pos}$. For $\Delta$t$_{\rm neg}$ the yield is almost constant irrespective of the pulse separation, being close to the rate \mbox{Y$^{\rm tot}_{\rm s}$} of the strong pulse alone. Pre-excitation with weak pulses instead results in a drop of the signal to about half of the former value Y$^{\rm tot}_{\rm s}$. Around zero delay the signal is strongly enhanced, partially originating from the temporal overlap of the pulses. However, the width of the feature is broader than expected from the autocorrelation (AC), see the discussion below and \mbox{Fig.~\ref{fig:fastdyn}} for a high resolution study. In addition, a minimum evolves at around $\Delta$t=+20\,ps. At larger optical delays the signal recovers, but at a lower level when compared to $\Delta$t$_{\rm neg}$.

\begin{figure}[t]
	\centering
	\vspace{0.5cm}
	\includegraphics[width=9cm]{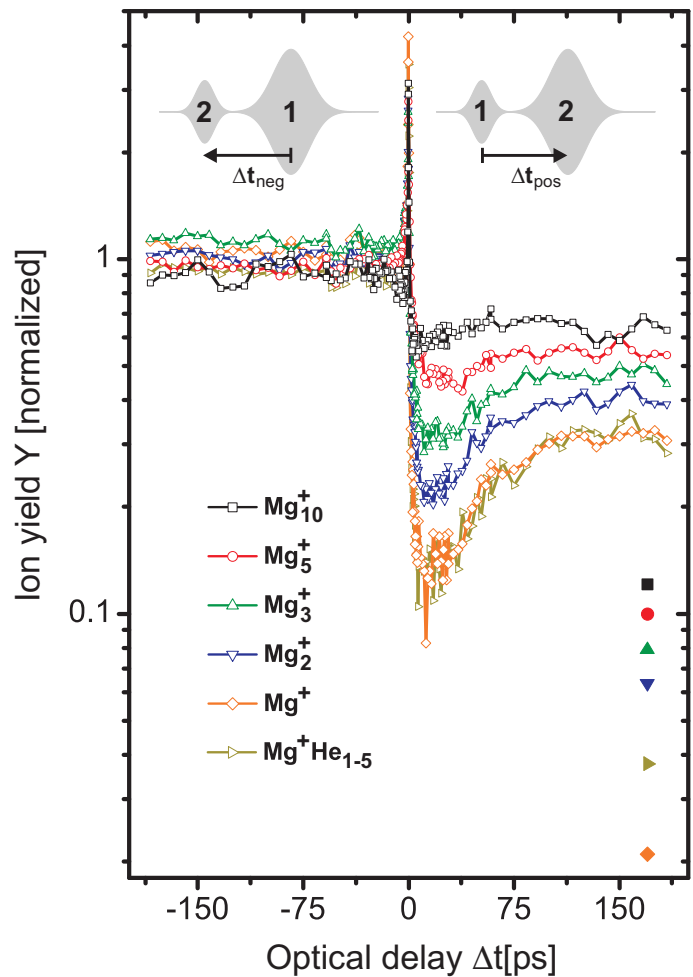}
	\caption{As \mbox{Fig.~\ref{fig:pptotalYield}}, but resolving for selected reaction products. The specific yields shown on a logarithmic scale are normalized to the signals after exclusively applying \emph{single} strong pulses. For $\Delta$t$_{\rm neg}$, ions are produced already in the interaction with the first pulse. The impact of the weaker trailing pulse is marginal. However, slight differences are obtained when comparing molecular ions (e.g. Mg$^+_3$) and clusters. Weak initial pulses instead ($\Delta$t$_{\rm pos}$) generate conditions where the impact of the strong pulse is reduced significantly. This holds true for all ion channels and is most obvious for Mg$^{+}$ and snowballs. In addition a strong dynamical feature is observed, which gets less pronounced for larger clusters. After the minima at about $\Delta$t=+20\,ps the signals recover on a timescale of 100\,ps. The values for \emph{single} weak pulse exposures are indicated (filled symbols).}
 \label{fig:ppClusteruFragments}
\end{figure}

%figure: selected ions
In order to analyse possible reaction pathways we now concentrate on
selected ionic species for the same dual-pulse sequence, see Fig.~\ref{fig:ppClusteruFragments}. The spectra are normalized to the yield of the single strong pulses. In general all interaction products qualitatively exhibit the signature of the total ion yield. Around zero delay a strong enhancement is observed for all ions, followed by a rapid drop within the first ps. At larger $\Delta$t$_{\rm neg}$ the second (weak) pulse contributes less and the yields stay close to that of the first strong pulse. However, the signals of larger clusters are slightly below those of smaller ones (e.g.\ Mg$^{+}_{3}$), irrespective of the delay. At positive delays $\Delta$t$_{\rm pos}$ all ions show a significantly reduced count rate, with specific responses to the delayed ionization. Over the entire timescale, the yields are lowest for atoms and snowballs. For clusters, the signal increases with N and is highest for Mg$^{+}_{\rm 10}$. However, the count rates stay a factor of two below the normalized single pulse reference. Furthermore, after the fast ps decay, Mg$^{+}_{\rm 10}$ shows a more or less time-independent signature while snowballs and small clusters depict strong dynamics in this region. The minima around $\Delta$t=+20\,ps are followed by a recovery of the signals on a timescale of about 100\,ps. We emphasize that especially around the dip no other ion channel gains intensity, which hints at a minimum in the ionization efficiency rather than controlling fragmentation pathways.

\section{Discussion}

\begin{figure}[t]
	\centering
	\vspace{0.5cm}
	\includegraphics[width=9cm]{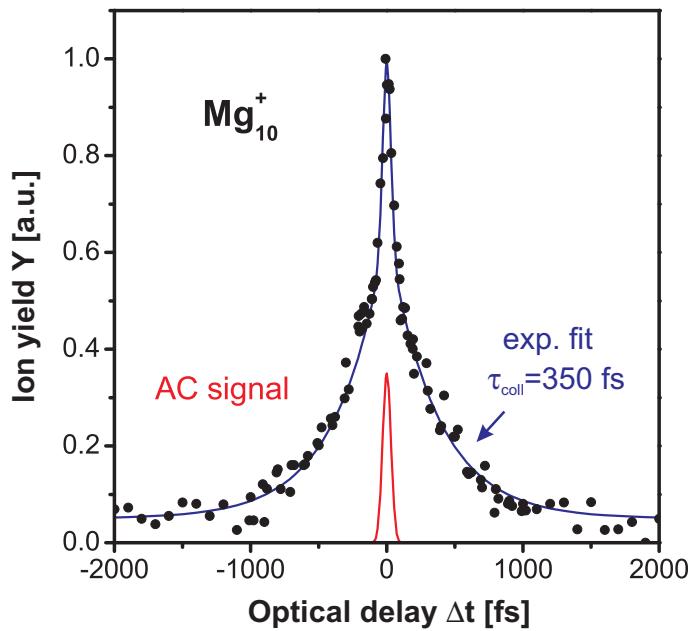}
	\caption{High resolution scan of Mg$^{+}_{10}$ close to zero optical delay with equal intensity dual pulses (60\,fs (FWHM), \mbox{$\rm I=8\times10^{11}\,W\,cm^{-2}$} each) using the technique of colored double
pulses~\cite{TruJPB11}. Mg$^+_{10}$ is chosen as representative for larger clusters. In addition to a fast correlation signal (red) the fit yields a decay time of \mbox{$\tau_{\rm coll}$=350\,fs} (blue) which we attribute to the foam collapse. All ionic channels show the same \mbox{$\tau_{\rm coll}$}.}
  \label{fig:fastdyn}
\end{figure}

A Mg foam in helium droplets would represent a unique state of
matter. Because the atoms form a regular 10 \AA-spaced network being
far off the closed packed cluster ground state, a substantial energy
reservoir is present. Perturbing the structure by, for example
photoabsorption will trigger a rapid collapse. The energy release
on the order of 0.5\,eV per atom will transiently convert the
condensate into a highly-excited system. Results from classical
molecular dynamics simulations, where the foam artificially undergoes a
sudden change to metallic binding, gives an implosion time of about one
ps~\cite{FenPC12} in rough accordance with the fast decay times obtained around zero delay, see \mbox{Fig.~\ref{fig:pptotalYield}} and
\mbox{Fig.~\ref{fig:ppClusteruFragments}}. In order to highlight this
time regime dual-pulse studies have been conducted with 60\,fs (FWHM) pulses of equal intensity. As an example for the foam response the signal of Mg$_{10}^+$ is shown in \mbox{Fig.~\ref{fig:fastdyn}}. The experimental data can be approximated by a two-term fit: an exponential decay of about \mbox{$\tau_{\rm coll}$=350\,fs} describing the collapse and an autocorrelation function which accounts for nonlinearities when the subpulses overlap. We like to emphasize, that similar \mbox{$\tau_{\rm coll}$} values are obtained for all ionic products.

In order to explain the fast decrease of the ion signal, we anticipate
that the collapse is reflected in a change of the ionization cross sections. With reduction of the interatomic distances during the collapse electronic properties are expected to evolve from atomic- to more bulk-like, which expresses the key concept of Mott~\cite{mott}. The transition is accompanied by an increase in level density which has an influence on the excited state lifetimes. In the metallic limit, lifetimes of excited states in clusters as short as fs have been determined~\cite{KnoSurScience96}. Thus the change in bond character and electron relaxation time is expected to reduce the MPI efficiency when compared
to the initial foam configuration.

At this point we like to mention the quantum optics aspect of the Mg
atom ensemble. In the network the large Mg-Mg distance causes a weak
overlap of the atomic wave functions, i.e. the foam represents a degenerate dipol ensemble which also introduces the question of collective excitation~\cite{MazJCP07}. For such systems, optical properties may be modified when compared to single atoms due to the quantum coherence of the atomic states. Hence the fast decay of the signal could be a signature of decoherence, i.e.\ the time for establishing disorder which goes along with the system collapse. For a theoretical description of this problem,
however, one would have to consider the coherent excitation of an
ensemble of atoms under multiphoton ionization conditions, a challenge
which, to our knowledge, has not been tackled so far.\\

Before discussing details of the collapse, an alternative scenario has to be considered, i.e.\ the response of initially compact clusters. Indeed,
depending on the system and on the cluster size typical electronic decay times are on the order of some 100\,fs, e.g.\ Ag$^{-}_{N}$~\cite{NiePRB07}, Pd$^{-}_{\rm N}$~\cite{PonJCP01}, Al$^{-}_{N}$~\cite{GerCPL03} and Au$^{-}_{7}$~\cite{StaPRA12}, which is close to our result. Also fragmentation can take place on the same timescale as observed for Na$_{\rm N}$~\cite{BauPRL92}. But all those studies exhibit strong size-dependent decay times in contrast to our measurements which gives a value of \mbox{$\tau_{\rm coll}$=350\,fs} for all products.  Hence, there is no evidence that the response of compact clusters is probed in our measurement. Moreover, with respect to the long-term dynamics the total yield should be related to dissociation. As depicted in Fig.~\ref{fig:pptotalYield}, a weak first pulse leads to a significant decrease in Y$^{\rm tot}$ compared to Y$^{\rm tot}_{\rm s}$. If compact clusters were present, any dissociation should increase the number of particles which can be probed by the subsequent ionizing pulse. Hence, a higher signal in Y$^{\rm tot}$ is expected for positive optical delays.  In contrast, our observation shows a significantly reduced Y$^{\rm tot}$ for $\Delta$t$_{\rm pos}$. This behavior is obtained irrespective of the chosen ion product, see \mbox{Fig.~\ref{fig:ppClusteruFragments}}.
Moreover, ionization of fragments should shift the size distribution
towards smaller N. Instead, the opposite trend is observed, e.g. in \mbox{Fig.~\ref{fig:ppClusteruFragments}} at positive delays the yield of larger clusters (Mg$_5^+$, Mg$_{10}^+$) is less reduced when compared to the smaller ones. Therefore, since neither a distinct size-dependence of \mbox{$\tau_{\rm coll}$} at short delays nor characteristic fragmentation signatures on ps timescales are observed, we conclude that compact clusters are unlikely to exist in the initially prepared system.\\

\begin{figure}[t]
	\centering
	\vspace{0.5cm}
	\includegraphics[width=15.0cm]{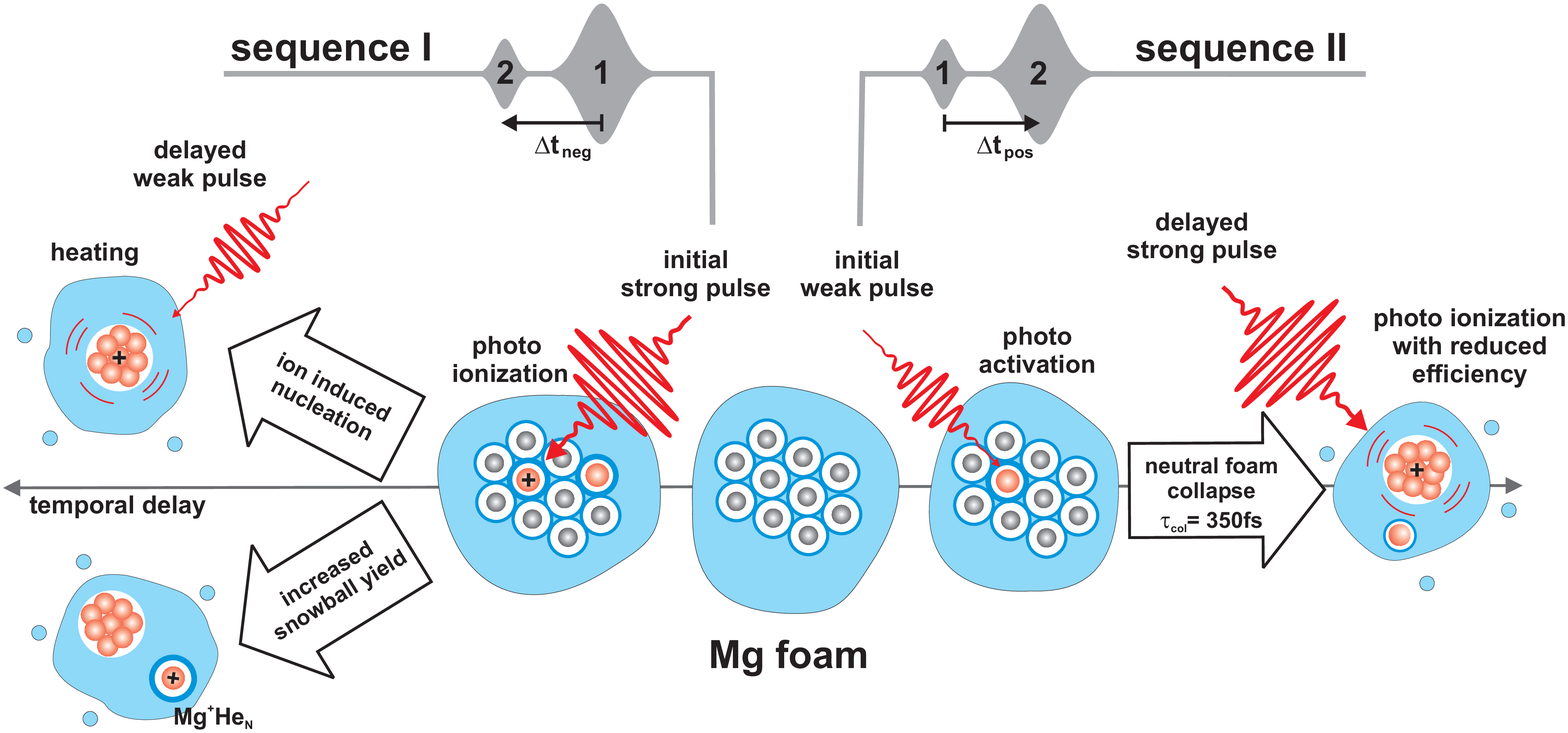}
	\vspace{0.5cm}
	\caption{Sketch of the Mg foam dynamics after dual-pulse excitation.
	Sequence (I)\,(middle to left): A strong initial pulse directly ionizes single atoms in the foam state. Ions act as seed forming charged compact
          clusters (\emph{ion induced nucleation}) and ion snowballs
          (\emph{enhanced snowball formation}). Weak
          trailing pulses do not further ionize but contribute with
          \emph{heating} of the ionic clusters. No delay dependence in the ion yields is expected. Sequence (II)\,(middle to right): A weak initial pulse photoactivates the foam and induces a fast relaxation into a
          neutral cluster (\emph{neutral foam collapse}). The strong
          trailing pulse hits on a compact system. Since the
          collapse is accompanied by a \emph{decrease in MPI efficiency}
          less ions are produced.}
\label{fig:foamDynamics}
\end{figure}

With respect to an induced collapse scenario it appears to be possible to rationalize the strong dependence of $\rm{Y}$ on the pulse order. For this we discuss two different decay pathways, i.e. (I) ion-induced nucleation for strong \mbox{($\Delta$t$_{\rm neg}$)} and (II) neutral foam dynamics for weak initial pulses \mbox{($\Delta$t$_{\rm pos}$)}. These sequences are illustrated in \mbox{Fig.~\ref{fig:foamDynamics}}.

In (I) strong  initial pulses create ionic seeds acting as nucleation centers for instantaneous formation of Mg$^{+}_{\rm N}$. As a competing channel, Mg$^+$-snowballs are produced favored by the enhanced helium density surrounding each impurity atom in the foam state. The weak trailing pulses have almost no influence on the charging. Instead, they may heat the nascent complexes and induce some fragmentation. Indeed, only a tiny enhancement in the signal of molecular ions (Mg$^{+}_{\rm 3}$) is obtained when compared to larger clusters, see the left side of \mbox{~\ref{fig:ppClusteruFragments}}. Moreover, a delay independent ion signal should evolve in agreement with our experimental observation. 

In sequence (II) the dynamics of the neutral photoactivated system is investigated. The weak preparatory pulse triggers a collapse of the neutral foam. The time evolution is probed by the delayed pulse which now ionizes the collapsing or the compact cluster. As discussed above, MPI of compact clusters should be reduced when compared to the initial foam. Thus, a strong dependence on the pulse order and a reduced $\rm{Y}$ for positive delays are expected, in accordance with the results in \mbox{Figs.~\ref{fig:pptotalYield} and ~\ref{fig:ppClusteruFragments}}. Note, that the formation of snowball complexes upon ionization of compact clusters is less likely compared to the foam state.

Induced by the weak pulse, the electronic configuration of embedded atoms can be changed leading to exciplex Mg$^*$He$_{\rm M}$ formation ~\cite{RehJCP00b,BruJCP01,SchPRL01}. The increased
local helium density around Mg$^*$ isolates the complex from the rest of
the foam. These systems are characterized by long-living excited
states and thus electronical relaxations can be excluded over the entire
timescale of the experiment. Compared to ions which tend to drift
towards the center of the droplets~\cite{LehMolPhys99,ZhaJCP12}, it is
well-known that excited atoms and exciplexes are expelled from the
host matrix, as a consequence of the extended valence orbital, see
e.g.~\cite{FedPRL99,LogJCP12}. Ionization within the droplet (i.e. short $\Delta$t) will lead to recombination into a cluster ion. In contrast, once an exciplex has left the droplet, cluster ions built by recombination are rather unlikely and snowballs are formed instead. Hence the signal on Mg$^+$He$_{\rm N}$ should show a time dependence which is directly related to the exciplex ejection. For droplet sizes of 10\,nm, ejection times on the order of 100\,ps are expected taking the Landau velocity (58m/s) into account. A corresponding signature is observed for \mbox{($\Delta$t$_{\rm pos}$)}, see
\mbox{Fig.~\ref{fig:ppClusteruFragments}}. The yields of atomic ions and snowballs show minima followed by a recovery of the signal. The increase of Mg$^+$ and Mg$^+$He$_{\rm M}$ can thus be interpreted as exciplex escape time which we estimate to about \mbox{$\tau_{\rm esc}$=50\,ps}. This consideration is supported by the fact that the minima are less developed for dimers, trimers and are almost absent for clusters.

\section{Summary and Conclusion}
By applying fs dual-pulse spectroscopy to a metastable Mg atom aggregate in He droplets, the collapse into a compact cluster has been studied. By applying pulses of a certain intensity ratio, the light-induced foam dynamics by either instantaneous ionization or neutral photoactivation can be addressed. On a picosecond timescale the yields exhibit a rapid drop. A decay time of about $\tau_{\rm coll}$=350\,fs has been determined, representing the collapse. The observed real-time dynamics touches fundamental issues, since the foam implosion might reflect the Mott transition. Further, the loss of the coherent character of the weakly interacting ensemble touches aspects of quantum optics. A qualitative model which accounts for most of the spectral features observed in the experiments supports our hypothesis from~\cite{PrzPRA08}, that a metastable aggregate has been identified in He droplets. In this work the focus was on excitation and charging dynamics at the lowest power to monitor exclusively dynamics in the multiphoton regime. At higher laser intensities, the onset of delayed plasmon enhanced multi-electron ionization~\cite{KoePRL99,FenRMP10} provides a another approach to probe the foam. Moreover these aggregates may serve as interesting targets for intense laser-cluster interaction studies.

\section{Acknowledgments}

We thank Stefan Scheel and Thomas Fennel for stimulating discussions. Further we are indepted to Peter Toennies and his group at the Max-Planck-Institute in G\"ottingen for providing us with the helium droplet technology. Financial support by the Deutsche Forschungsgemeinschaft through SFB652 is acknowledged.
%\bibliographystyle{unsrt}
%\bibliography{GoeNJP12}

\end{document}